\documentclass[12pt]{article}\usepackage{EntBellMix}

\begin{document}

\title{Entanglement of Bell Mixtures\\
of Two Qubits}

\author{Robert R. Tucci\\
        P.O. Box 226\\ 
        Bedford,  MA   01730\\
        tucci@ar-tiste.com}

\date{ \today} 

\maketitle

\vskip2cm
\section*{Abstract}
This paper is an appendix to a previous paper:
quant-ph/0101123 
``Relaxation Method for Calculating Quantum Entanglement", by Robert Tucci.
For certain mixtures of Bell basis states, namely the Werner States, we 
use the theoretical machinery of our previous paper to derive
algebraic formulas for: 
the pure and mixed minimization entanglements 
(i.e., $E_{pure}$ and $E_{mixed}$),
their optimal decompositions and 
their entanglement operators.
This complements and corroborates some results that
were obtained numerically but not algebraically in 
our previous paper. 
Some of the algebraic formulas presented here are new. Others
were first derived using a different method by 
Bennett et al in quant-ph/9604024.

\newpage
\section{Introduction}

This paper is an appendix to a previous paper\cite{Tuc}
by the same author.
We will assume that the reader has read our previous paper.
Without having done so, he/she won't be able to 
understand this paper beyond its Introduction.

Henceforth, we will use ``min." as an abbreviation for
the word ``minimization".
In our previous paper\cite{Tuc}, we defined two quantum entanglement
measures, the {\it pure min. entanglement} ($E_{pure}$) and 
the {\it mixed min. entanglement} ($E_{mixed}$).
These measures apply to any bipartite density matrix
(the subscripts refer to the type of minimization space used,
not to whether the density matrix is pure or mixed.)
We showed that $E_{pure}$ is equal to
the entanglement of formation, a measure of
entanglement first defined by Bennett et al in Ref.\cite{Ben}. 
$E_{mixed}$, on the other hand, is a 
new animal.  We gave a numerical method 
for calculating $E_{pure}$ and $E_{mixed}$,
their optimal decompositions, and also their
entanglement operators (operators
whose expectation value gives the entanglement). 
We gave numerical results obtained with Causa Com\'{u}n, a computer 
program that implements the ideas of Ref.\cite{Tuc}.
We did this for a special type of Bell mixture called a Werner State and
for Horodecki States that exhibit bound entanglement.

In Ref.\cite{Ben}, Bennett et al derived an 
explicit algebraic formula for the entanglement of formation
of any Bell mixture. In Ref.\cite{Woo}, Wootters went 
one step further and generalized
the formula of Ref.\cite{Ben} to encompass all density matrices of two qubits. 

In this paper, we 
use the theoretical machinery of our previous paper to derive
certain algebraic formulas for Werner States.
Specifically, we give explicit algebraic formulas 
for
$E_{pure}$ and $E_{mixed}$,
their optimal decompositions and 
their entanglement operators.
This complements and corroborates some results that
were obtained numerically but not algebraically in 
our previous paper. 
Most of our formulas  for $E_{pure}$
were first derived, using a different method, by
Bennett et al in Ref.\cite{Ben}.
Our formulas for $E_{mixed}$ are new.

\section{Notation}

We assume the reader is familiar with the notation of Ref.\cite{Tuc}.
In this section we will introduce some additional notation
that  is used throughout this paper. 

We will use the notation of Ref.\cite{Tuc}
intact except for one small modification.  
Ref.\cite{Tuc} dealt with a Hilbert space $\hil_{\rvx\rvy} = \hil_\rvx\otimes \hil_\rvy$.
Its two parts  were represented by
the random variables $\rvx$ and $\rvy$ (Xerxes and Yolanda). Here we
will rename the two parts $\rva$, $\rvb$ (Alice and Bob). This 
conforms more closely with the rest of the literature.
Also, it looks better in cases such as the one
considered in this paper where one also uses $x,y,z$ for indices 
of Pauli matrices. In conclusion, throughout this paper, we will be dealing with 
$\hilab$ where $S_\rva = S_\rvb = Bool$.

For any Hilbert space $\hil$ and any $\ket{\psi}\in \hil$, we will often represent 
the projection operator $\ket{\psi}\bra{\psi}$ by $\pi(\psi)$. 
${\cal L}(\hil)$ will denote the set of linear operators acting on $\hil$.

Let $Z_{j, k}$ be the set of integers from $j$ to $k$,
including both $j$ and $k$. Let $Bool = \{0, 1\}$.
Let $x^{\#n}$ be the $n$-tuple with $x$ repeated $n$ times. For example,
$0^{\#3} = 0,0,0$.

The 
Kronecker delta function $\delta(x,y)$ equals one if $x=y$ and zero otherwise.
We will often abbreviate $\delta(x,y)$ by $\delta^x_y$, 
$\delta(x,y)\delta(p,q)$ by $\delta^{xp}_{yq}$, etc.
Also, we will use $\delta^{\no{\nu}}_\mu$ as an
abbreviation for $1-\delta^\nu_\mu$. In other words, 
$\delta^\no{\nu}_\mu$ is an 
an indicator function which equals 1 whenever $\mu\neq \nu$ and 
zero when $\mu= \nu$. For example, if $\mu, \nu\in Z_{0,3}$,
then the metric in Special Relativity can be written as 
$g_{\mu\nu} = (\delta_{\mu\nu}^{00} - \delta_{\mu\nu}^{\no{0}\no{0}} )\delta^\mu_\nu$.
One must be careful not to confuse 
$\delta_{\mu\nu}^{\no{\alpha}\no{\beta}}=
(1-\delta_\mu^\alpha)(1-\delta_\nu^\beta)$
and $\delta_{\mu\nu}^{\no{\alpha\beta}}=
1 - \delta_\mu^\alpha\delta_\nu^\beta$.

We will often use the {\it color summation convention}\cite{color-sum}. By this we 
mean that the summation signs will not be shown; summation
will instead be indicated by displaying summed indices in a different 
color than the unsummed ones. For example, $F_{\mu\sumnu}v^\sumnu = \sum_\nu F_{\mu\nu}v^\nu$.
This is a better notation than the {\it Einstein 
implicit summation convention} which it is meant to replace. 
In the Einstein convention, we are instructed to sum over repeated indices.
This becomes clumsy and requires a warning to the reader 
 whenever  we wish to use
repeated indices that are not summed over.

As is common in Relativity texts, we will often use Greek letters to represent 
indices that range over $Z_{0,3}$ and Latin letters to represent 
indices that range over $Z_{1,3}$. Unlike Relativity texts, we will not
distinguish between upper and lower indices.

Define $\invtwo = diag(1, -1, 1)$.
For any 3-dimensional vector $\vec{n} = (n_1, n_2, n_3)^T$, 
$\invtwo\vec{n} = (n_1, -n_2, n_3)^T$. One can 
likewise define ${\cal F}_j$ for  $j\in Z_{1,3}$ to be an operator that  
``flips" the $j$th component of the vector it acts on.

For any 3-dimensional vectors $\vec{a}$ and $\vec{x}$, 
$\vec{a}\times \vec{x} = A\vec{x}$, where

\beq
A = \left[
\begin{array}{ccc}
0 & -a_3 & a_2 \\
a_3 & 0 & -a_1 \\
-a_2 & a_1 & 0 
\end{array}
\right]
\;.
\eeq
We will often represent $A$ by $[\vec{a} \times \cdot]$.

Let

\beq
\ket{0}=
\left(
\begin{array}{c}
1\\0
\end{array}
\right)
\;,\;\;
\ket{1}=
\left(
\begin{array}{c}
0\\1
\end{array}
\right)
\;.
\eeq
Let $\sigma^0=1$. Let $\vec{\sigma}$ be the 3 dimensional vector of Pauli matrices.
The Pauli matrices are defined by

\beq
\sigma_x =
\left(
\begin{array}{cc}
0&1\\
1&0
\end{array}
\right)
\;,\;\;
\sigma_y =
\left(
\begin{array}{cc}
0&-i\\
i&0
\end{array}
\right)
\;,\;\;
\sigma_z =
\left(
\begin{array}{cc}
1&0\\
0&-1
\end{array}
\right)
\;.
\eeq
As is well known, the Pauli matrices satisfy:

\beq
\sigma^k \sigma^r = \delta^r_k + i \epsilon_{kr\sumj}\sigma^\sumj
\;,
\label{eq:sigma-prod}\eeq
for $k,r,j\in Z_{1,3}$, where $\epsilon_{krj}$ is the totally antisymmetric tensor with 
$\epsilon_{123} = 1$. Unfortunately, there is no formula 
for $\sigma^\mu \sigma^\nu$ with  $\mu, \nu\in Z_{0,3}$, that
matches the conciseness and standardization of Eq.(\ref{eq:sigma-prod}).
Here is one particular attempt.

\beq
\sigma^\mu \sigma^\nu = f_{\mu \nu} \sigma^{\mu\oplus\nu}
\;,
\eeq
where

\beq
\mu\oplus\nu =
\begin{tabular}{r|rrrrr}
& {\tiny 0} & {\tiny 1}& {\tiny 2}& {\tiny 3} & {\small $\nu$}\\
\hline
{\tiny 0}& 0& 1& 2& 3&\\
{\tiny 1}& 1& 0& 3& 2&\\
{\tiny 2}& 2& 3& 0& 1&\\
{\tiny 3}& 3& 2& 1& 0&\\
{\small $\mu$}& & & & &
\end{tabular}
\;,
\label{eq:group-table}\eeq
and

\beq
f_{\mu\nu} =
\begin{tabular}{r|rrrrr}
& {\tiny 0} & {\tiny 1}& {\tiny 2}& {\tiny 3} & {\small $\nu$}\\
\hline
{\tiny 0}& 1& 1& 1& 1&\\
{\tiny 1}& 1& 1& $i$&-$i$&\\
{\tiny 2}& 1&-$i$& 1& $i$&\\
{\tiny 3}& 1& $i$&-$i$& 1&\\
{\small $\mu$}& & & & &
\end{tabular}
\;.
\label{eq:fmunu-table}\eeq

Note that the operation $\mu\oplus\nu$ defined by Eq.(\ref{eq:group-table})
specifies an Abelian group (the operation is 
commutative, associative, has an identity, and has
an inverse for each of its elements). The Abelian group defined by  $\oplus$ on $Z_{0,3}$
can be shown to be simply the product of two copies of the group of two elements.

Instead of defining $f_{\mu\nu}$ by the
table Eq.(\ref{eq:fmunu-table}), one can define it by the rather clumsy expression:

\beq
f_{\mu\nu} = 
\delta^\nu_\mu + 
\delta^{\no{0} 0}_{\mu\nu} +
\delta^{0 \no{0}}_{\mu\nu} +
\delta^{\no{0}\no{0}}_{\mu\nu} i\epsilon_{\mu,\nu,\mu\oplus\nu} 
\;.
\eeq
$f_{\mu\nu}$ has a few useful properties. For example, it is Hermitian and it satisfies:

\beq
f^*_{\alpha\oplus\beta, \beta} = f_{\alpha, \beta}
\;.
\eeq

For any $x\in[0, 1]$, the {\it binary entropy function} $h(x)$ is defined by
\beq
h(x) = -x \log_2 (x) - (1-x) \log_2 (1-x)
\;.
\eeq
Occasionally, we will also need to use $h(x)$ with the base 
2 logs replaced by base $e$ ones. 
So define

\beq
h_e(x) = (\ln 2)h(x) =  -x \ln(x) - (1-x) \ln (1-x)
\;.
\eeq

\section{Bell Basis} \label{sec:bell-basis} 
In this section we will discuss various properties of the Bell Basis.

One can define operators that act only on the $\hil_\rva$ (ditto, 
$\hil_\rvb$) part of $\hilab$. Let $\siga{\mu}$ (ditto, $\sigb{\mu}$) for $\mu\in Z_{0,3}$
represent the Pauli matrices that act on space $\hil_\rva$ (ditto,
$\hil_\rvb$). Another natural notation
for these operators is $\sigma_\rva^\mu$ and $\sigma_\rvb^\mu$.

The following four states are usually called the ``Bell basis" 
(with the magic phases) of $\hilab$:
\beq
\ketbell{0} 
=\ket{=^+} 
= \frac{1}{\sqrt{2}}(\ket{00} + \ket{11})
\;,
\eeq

\beq
\ketbell{1} 
=i\sigb{1}\ket{=^+}
=\frac{i}{\sqrt{2}}(\ket{01} + \ket{10})
= i\ket{\neq^+}
\;,
\eeq

\beq
\ketbell{2} 
=i\sigb{2}\ket{=^+}
=\frac{-1}{\sqrt{2}}(\ket{01} - \ket{10})
= -\ket{\neq^-}
\;,
\eeq

\beq
\ketbell{3} 
=i\sigb{3}\ket{=^+}
=\frac{i}{\sqrt{2}}(\ket{00} - \ket{11})
= i\ket{=^-}
\;.
\eeq
(By taking matrix products and linear combinations with {\bf real} coefficients, of the
operators 1, $i\sigma_x$, $i\sigma_y$ and $i\sigma_z$,
one generates what is called the Quaternion Algebra, invented by Hamilton.)

The Bell basis states are an orthonormal basis of $\hilab$ so they satisfy

\beq
\av{ B(\mu) | B(\nu) } = \delta^\mu_\nu
\;,
\eeq
and

\beq
\ketbell{\summu}\brabell{\summu}=1
\;.
\eeq

The Bell basis states place listeners $\rva$ and $\rvb$ 
on equal footing: measurement of $\siga{\mu}$ 
is the same as measurement of $\sigb{\mu}$ up to a sign. Indeed,
the action of $\siga{\mu}$ on $\ketbell{0}$ is
\beq
\siga{\mu}\ket{=^+} = (-1)^{\delta^2_\mu}\sigb{\mu}\ket{=^+}
\;.
\eeq
The action of
$\siga{\mu}$ on $\ketbell{\nu}$ for  $\nu \neq 0$
may have an additional $-1$ factor due to 
the fact that the Pauli matrices anticommute.
For example,
\beqa
\siga{2}\ketbell{3} 
&=&
\siga{2} i\sigb{3}  \ket{=^+}
=i\sigb{3} \siga{2} \ket{=^+}
=-i\sigb{3} \sigb{2}\ket{=^+} =
\nonumber\\
&=& \sigb{2} i \sigb{3}\ket{=^+} = \sigb{2}\ketbell{3} 
\;.
\eeqa
Thus, we see that, in general, the action of
$\siga{\mu}$ on $\ketbell{\nu}$ is
\beq
\siga{\mu}\ketbell{\nu} 
= (-1)^{\delta^2_\mu}
(-1)^{\delta^{\no{0}\no{0}}_{\mu \nu}\delta^\no{\mu}_\nu}
\sigb{\mu}\ketbell{\nu}
\;.
\eeq

Suppose $\Omega_\rvb$ (ditto, $\Omega_\rva$)
is an operator acting only on $\hilb$ (ditto, $\hila$). 
We will call such operators {\it local} operators.
It is easy to show that
\beq
\brabell{\mu} \Omega_\rvb \ketbell{\mu} 
= \av{=^+ | \Omega_\rvb | =^+} 
= \frac{1}{2} \tr(\Omega_\rvb)
\;,
\eeq
and
\beq
\brabell{\mu} \Omega_\rva \ketbell{\mu} 
=\av{=^+ | \Omega_\rva | =^+} 
= \frac{1}{2} \tr(\Omega_\rva)
\;.
\eeq
Note the right hand sides are independent of $\mu$ 
(although $\brabell{\mu} \Omega_\rvb \ketbell{\nu}$ 
generally does depend on $\mu$ and $\nu$.)
Hence, all 4 Bell basis states
contain the same amount of information about  
expected values of local operators.

In future sections, we will need to find the matrix elements in the Bell 
basis of certain operators in ${\cal L}(\hilab)$. These operators can 
always be written as a linear combination $x_{\summu\sumnu}\siga{\summu}\sigb{\sumnu}$.
In this linear combination, $\siga{\mu}$ will be acting on a Bell state so it
can be replaced by plus or minus $\sigb{\mu}$. The product $\sigb{\mu}\sigb{\nu}$ can itself
be replaced by $f_{\mu\nu} \sigb{\mu\oplus\nu}$. 
In this way, we can reduced the problem of calculating
the matrix elements in the Bell 
basis of any operator in ${\cal L}(\hilab)$ to
calculating the matrix elements in the Bell 
basis of $\sigb{\beta}$. One has:  
\beqa
\brabell{\mu_1} \sigb{\beta} \ketbell{\mu_2} 
&=&
(-i)^{\delta^\no{0}_{\mu_1}}
f_{\mu_1, \beta\oplus\mu_2}
f_{\beta, \mu_2}
(i)^{\delta^\no{0}_{\mu_2}}
\av{=^+|\sigb{\beta\oplus\mu_1\oplus\mu2}|=^+} 
\nonumber\\
&=&
(-i)^{\delta^\no{0}_{\mu_1}}
f_{\beta, \mu_2}
(i)^{\delta^\no{0}_{\mu_2}}
\delta_{\beta\oplus\mu_1\oplus\mu2}^0
\;.
\eeqa
An equivalent way of writing the last equation is:
\beq
\brabell{\mu_1} \sigb{0} \ketbell{\mu_2} =\delta^{\mu_1}_{\mu_2}
\;,
\eeq
and

\beq
\brabell{\mu_1} \vec{x}\cdot\vecsigb \ketbell{\mu_2}=
\left[
\begin{array}{cc}
0 & i \vec{x}^T \\
-i\vec{x} & i(\vec{x}\times\cdot)
\end{array}
\right]_{\mu_1, \mu_2}
\;.
\eeq

Another problem that we shall encounter in future sections
is finding the partial trace with respect to either $\rva$ or
$\rvb$ of an operator $X \in {\cal L}(\hilab)$. If  
\beq
X = x_{\summu\sumnu} \ketbell{\summu}\brabell{\sumnu}
\;,
\eeq
then one finds that

\beq
\tr_\rva X = 
\frac{1}{2}
\left\{
x_{\summu\summu} 
+ \left[
x_{\sumk 0} -x_{0 \sumk}
+ x_{\sump\sumq}
\epsilon_{\sump\sumq\sumk}
\right]
i \sigb{\sumk}
\right\}
\;,
\eeq
and

\beq
\tr_\rvb X = 
\frac{1}{2}
\left\{
y_{\summu\summu} 
+ \left[
y_{\sumk 0} -y_{0 \sumk}
+ y_{\sump\sumq}
\epsilon_{\sump\sumq\sumk}
\right]
i \siga{\sumk}
\right\}
\;,
\eeq
where $y_{\mu\nu} = x_{\mu\nu}(-1)^{\delta_\mu^2}(-1)^{\delta_\nu^2}$.
These equations generalize the well known result:

\begin{subequations}
\beq
\tr_\rva \ketbell{\mu}\brabell{\mu} = \frac{1}{2}\left(\pi(\ket{0}_\rvb) +  \pi(\ket{1}_\rvb)\right)
\;,
\eeq

\beq
\tr_\rvb \ketbell{\mu}\brabell{\mu} = \frac{1}{2}\left(\pi(\ket{0}_\rva) +  \pi(\ket{1}_\rva)\right) 
\;.
\eeq
\end{subequations}
Thus, the Bell basis states do not distinguish between 
$\ket{0}_\rva$ and $\ket{1}_\rva$ (ditto, $\ket{0}_\rvb$ and $\ket{1}_\rvb$)
when the state of $\rvb$ (ditto, $\rva$) is unknown.
An immediate consequence of the last equation is that
the Bell states have maximum entanglement of formation. 

An element of  
${\cal L}(\hilab)$ can be expanded
in various bases: one can expand it in terms 
of the operators $\siga{\mu}\sigb{\nu}$ for all $\mu, \nu\in Z_{0,3}$ (call this the Pauli ${\cal L}(\hilab)$-basis),
or the operators $\ket{a,b}\bra{a',b'}$ for all $a,b,a', b'\in Bool$ (call this the Standard ${\cal L}(\hilab)$-basis),
or the operators $\ketbell{\mu}\brabell{\nu}$ for all $\mu, \nu \in Z_{0,3}$ (call this the Bell ${\cal L}(\hilab)$-basis).
In what follows, we will use mostly the Bell ${\cal L}(\hilab)$-basis because 
it seems the most natural one for doing entanglement calculations.
Thus, henceforth, whenever we represent ${\cal L}(\hilab)$ operators by 4 by 4 matrices,
the matrices should be understood as representations in the Bell ${\cal L}(\hilab)$-basis.

\section{Entanglement of Pure State}
In this section we calculate the entanglement of 
any pure state of two qubits\cite{Ben}. This is a good
warm up exercise to prepare us for the following sections, 
where we address the harder problem of calculating entanglements of 
mixed states. 

Below, for any complex
vector $\vec{z}$, we will use
$|\vec{z}| =\sqrt{\vec{z}\cdot\vec{z}^{\;*}}$ and  $\vec{z}^{\;2} = \vec{z}\cdot\vec{z}$.

Any unit length $\ket{\psi}\in\hilab$ can be expressed 
in the Bell basis as:
\beq
\ket{\psi} = (z_0 + i\vec{z}\cdot \vecsigb)\ket{=^+}
\;,
\eeq
where $\av{\psi|\psi} = |z_0|^2 + |\vec{z}|^2 = 1$.  
If

\beq
\rho = \ket{\psi}\bra{\psi} =
(z_0 + i\vec{z}\cdot \vecsigb)
\ket{=^+}\bra{=^+}
(z^*_0 -i \vec{z}^{\;*}\cdot \vecsigb)
\;,
\eeq
then

\beq
\tr_\rva \rho = 
\frac{1}{2}(z_0 + i\vec{z}\cdot \vecsigb)
(z^*_0 - i\vec{z}^{\;*}\cdot \vecsigb)
=
n_0 + \vec{n}\cdot \vecsigb
\;,
\label{eq:tr_a_rho}\eeq
where

\beq
n_0 = \frac{1}{2}
\;,
\eeq
and

\beq
\vec{n} =
\frac{i}{2}
( z^*_0 \vec{z} - z_0\vec{z}^{\;*} + \vec{z} \times \vec{z}^{\;*} )
\;.
\eeq

From Eq.(\ref{eq:tr_a_rho}) and Appendix \ref{app:eigensystem}, the eigenvalues of
$\tr_\rva \rho$ are simply $n_0 \pm |\vec{n}|$. Hence,

\beq
E_{pure} = E_{mixed} = h(n_0 + |\vec{n}|)
\;.
\eeq
One can show using well known vector product identities that 
for any 4-tuple $(z_0, \vec{z})$ of complex numbers such that
$|z_0|^2 + |\vec{z}|^2 = 1$, one has

\beq
|z^*_0 \vec{z} - z_0\vec{z}^{\;*} + \vec{z} \times \vec{z}^{\;*}|^2 = 
1 - |z_0^2 + \vec{z}^{\;2}|^2
\;.
\eeq
Hence

\beq
E_{pure} = E_{mixed} = h(\frac{1 + \sqrt{ 1 - C^2}}{2})
\;,
\eeq
where

\beq
C = |z_0^2 + \vec{z}^{\;2}|
\;.
\eeq

$C$ is called the {\it concurrence}\cite{Ben} of $\ket{\psi}$.
$E_{pure}$ is a monotonically nondecreasing function of $C$, 
and they both vanish at the same time, so $C$ is 
also a good measure of entanglement. $0\leq C \leq 1$.
The pure state $\ket{\psi}$ has  $C=1$ (maximum entanglement) iff 
its coefficients $(z_0, \vec{z})$ are all real.

\section{Entanglement of Bell Mixture}\label{sec:main}
In this section we present the main calculation of this paper.
For Werner states, we calculate $E_{pure}$ and $E_{mixed}$, and  
their corresponding optimal decompositions and entanglement operators.
Our calculation is split into 4 parts: 
(1)$\alpab{K}$, (2)$\alpab{R}$ 
(3)$E_{pure}$ and $E_{mixed}$
(4)$\Delta_\rvab$.

We will call 
a {\it Bell mixture} any density matrix $\rho_\rvab$ that can be expressed as

\beq
\rho_\rvab = \sum_\mu m_\mu \ketbell{\mu}\brabell{\mu}
\;,
\eeq
where $\sum_\mu m_\mu=1$.
We will call a {\it Werner state} any state that can be expressed as

\beq
\rho_\rvab = m_0 \ketbell{0}\brabell{0} + m_1 \sum^\dimv_{\mu= 1} \ketbell{\mu}\brabell{\mu}
\;,
\eeq
where $\dimv\in\{1,2,3\}$ and $m_0 + \dimv m_1 = 1$. This is a slight generalization
from what is commonly called a Werner state. The term Werner state 
usually refers to the case where $\dimv=3$ and the rank of $\rho_\rvab$ is 4.

\subsection{$\alpab{K}$ Calculations} \label{sec:k}

We will begin by assuming  that
$\alpab{K}$ can be expressed 
in a special ``ansatz" form.

Define

\beq
I_v = diag(1^{\#\dimv}, 0^{\#3-\dimv})
\;,
\eeq
where $\dimv\in Z_{1,3}$ was defined previously.
Now  assume $\alpab{K}$ can be expressed in the Bell representation as:

\beq	
\alpab{K} = 
\frac{1}{N_\rvalp}
\left[
\begin{array}{ll}
m_0 & iq \vecalpt{v} \\
-iq \vecalp{v} & m_1 \vecalp{v}\vecalpt{v} + \epsilon m_1 (I_v - \vecalp{v}\vecalpt{v}) 
\end{array}
\right]
\;.
\label{eq:k-candidate}\eeq
We assume that $q$ and $\epsilon$ are real, 
$m_0 + \dimv m_1 = 1$, and $\{\vecalp{v} | \alpha\in Z_{1, N\rvalp}\}$ is a set of real
3-dimensional vectors satisfying

\begin{subequations}

\beq
\vecalpt{v}\vecalp{v} = \dimv
\;
\eeq
for all $\alpha\in Z_{1, N\rvalp}$,
 
\beq
\sum_{\alpha=1}^{N_\rvalp} \vecalp{v} = 0
\;,
\eeq
and

\beq
\sum_{\alpha=1}^{N_\rvalp} \vecalp{v}\vecalpt{v} = N_\rvalp I_v
\;.
\eeq
\label{eq:v-constraints}\end{subequations}
Here are some examples of sets of $\vecalp{v}$'s that satisfy Eqs.(\ref{eq:v-constraints}).

\begin{subequations}
\beq 
\begin{array}{ll}
 \dimv = 1,\; N_\rvalp =2, & 
\vecalp{v} \in 
\left \{
\left( \begin{array}{c} 1\\0\\0 \end{array} \right),\;
\left( \begin{array}{c} -1\\0\\0 \end{array} \right)
\right \}
\end{array}
\;,
\eeq

\beq 
\begin{array}{ll}
\dimv = 2,\; N_\rvalp =4,  & 
\vecalp{v} \in 
\left \{
\left( \begin{array}{c} (-1)^a\\(-1)^b\\0 \end{array} \right)
| a, b \in Bool 
\right \}
\end{array}
\;,
\eeq

\beq 
\begin{array}{ll}
\dimv = 3,\;  N_\rvalp =4, & 
\vecalp{v} \in 
\left \{
\left( \begin{array}{c}  1\\ 1\\ 1 \end{array} \right),\;
\left( \begin{array}{c}  1\\-1\\-1 \end{array} \right),\;
\left( \begin{array}{c} -1\\ 1\\-1 \end{array} \right),\;
\left( \begin{array}{c} -1\\-1\\ 1 \end{array} \right)
\right \}
\end{array}
\;,
\eeq

\beq 
\begin{array}{ll}
\dimv = 3,\; N_\rvalp =8, & 
\vecalp{v} \in 
\left \{
\left( \begin{array}{c} (-1)^a\\(-1)^b\\(-1)^c \end{array} \right)
| a, b, c \in Bool 
\right \}
\end{array}
\;.
\eeq
\label{eq:v-alp-egs}
\end{subequations}

Note that
\beq
\rho_\rvab = \sum_\alpha \alpab{K} = diag(
m_0, 
(m_1)^{\#\dimv},
0^{\#3-\dimv}
)
\;,
\eeq
and 

\beq
w_\alpha = \tr_\rvab \alpab{K} = 1/N_\rvalp
\;.
\eeq
For $E_{pure}$, it is clear that
we want $q=\sqrt{m_0 m_1}$ and $\epsilon=0$ in Eq.(\ref{eq:k-candidate}).
For $E_{mixed}$, we intend to find those values of 
$q$ and $\epsilon$ that give the smallest possible 
conditional mutual information.

Our ansatz $\alpab{K}$ given by Eq.(\ref{eq:k-candidate}) 
depends on the following parameters:
$m_0$, $m_1$, $\dimv$, $N_\rvalp$, $q$, $\epsilon$ and the $\vecalp{v}$.
Out of these primitive parameters, one can 
construct the following auxiliary parameters
whose use will significantly shorten our subsequent 
formulas.

\begin{subequations}
\beq
\eta = \dimv - \epsilon(\dimv - 1)
\;,
\eeq

\beq
u = m_0 + \eta m_1
\;,
\eeq

\beq
k = \frac{ m_0 - \eta m_1 }{2} 
\;,
\eeq

\beq
\yy = |q| \sqrt{\dimv}
\;,
\eeq

\beq
\xx = \sqrt{ k^2 + \yy^2}
\;.
\eeq
\label{eq:aux-params}\end{subequations}

Next we will find the eigenvalues of $\alpab{K}$.
This can be done by using the following well known formula.
Suppose $M$ is a square matrix that can be partitioned
into 4 blocks $A, B, R_1, R_2$:

\beq
M = 
\left[
\begin{array}{cc}
A & R_2 \\
R_1 & B
\;
\end{array}
\right ]
\;,
\eeq
where the submatrices $A$ and $B$ are square but
$R_1$ and $R_2$ need not be. Then one can show that

\beq
\det(M) = \det(A) \det( B - R_1 \frac{1}{A} R_2)
\;.
\eeq
One can use the last equation to find
the eigenvalues of our ansatz $\alpab{K}$. One finds (independent of $\alpha$):

\beq
\begin{array}{|l|l|}
\hline
eigenvalue & degeneracy\\
\hline\hline
\lambda_+ = \left( \frac{u}{2} + \xx \right)/N_\rvalp & 1 \\
\hline
\lambda_- = \left( \frac{u}{2} - \xx \right)/N_\rvalp & 1 \\
\hline
\lambda_0 = \epsilon m_1/N_\rvalp & \dimv - 1 \\
\hline
0 & 3 - \dimv\\
\hline
\end{array}
\;.
\eeq

Note that since the eigenvalues of $\alpab{K}$ must be non-negative,
we must have $\epsilon\geq 0$ and $\frac{u}{2} - \xx \geq 0$. Since

\beq
(\frac{u}{2} - \xx)(\frac{u}{2} + \xx) 
= (\frac{u}{2})^2 - \xx^2 = 
m_0 m_1\eta - q^2\dimv
\;,
\eeq
it follows that $|q|\leq \sqrt{m_0 m_1 \frac{\eta}{\dimv}} \leq \sqrt{m_0 m_1}$.
One can also assume without loss of generality that $q\geq 0$ since if $q<0$, then one
can replace $q\rarrow - q$ and $\vecalp{v}\rarrow - \vecalp{v}$ for all $\alpha$.

We also need to know $\ln \alpab{K}$. To 
calculate $\ln \alpab{K}$,
it is not enough to find the eigenvalues of $\alpab{K}$;
we also need to find its eigenvectors.
Our technique for finding them is inspired by  
Appendix \ref{app:eigensystem}, where we find
the eigensystem 
of any 2 by 2 Hermitian matrix. 

We begin by defining, for each $\alpha$,
three operators called
$E_\alpha$, $\Sigma_\alpha$ and $P^{(0)}_\alpha$:

\beq
E_\alpha = 
\left[
\begin{array}{ll}
1 & 0 \\
0 & \frac{\vecalp{v}\vecalpt{v}}{\dimv}
\end{array}
\right]
\;,
\eeq

\beq
\Sigma_\alpha = 
\frac{1}{\xx}
\left[
\begin{array}{ll}
k & iq\vecalpt{v} \\
-iq\vecalp{v} & -k\frac{\vecalp{v}\vecalpt{v}}{\dimv}
\end{array}
\right]
\;,
\eeq

\beq
P^{(0)}_\alpha = 
\left[
\begin{array}{ll}
0 & 0 \\
0 & I_v - \frac{\vecalp{v}\vecalpt{v}}{\dimv}
\end{array}
\right]
\;.
\eeq
Note that these operators satisfy the following
multiplication table:

\beq
\begin{array}{l|ccc}
& E_\alpha & \Sigma_\alpha & P^{(0)}_\alpha\\
\hline
E_\alpha & E_\alpha & \Sigma_\alpha & 0\\
\Sigma_\alpha & \Sigma_\alpha & E_\alpha & 0\\
P^{(0)}_\alpha & 0 & 0 & P^{(0)}_\alpha
\end{array}
\;.
\eeq
$E_\alpha$ and $\Sigma_\alpha$ can be used to define two new operators 
$P^{(\pm)}_\alpha$:  

\beq
P^{(\pm)}_\alpha = \frac{ E_\alpha \pm \Sigma_\alpha}{2}
\;.
\eeq
Note that the $P^{(\sigma)}_\alpha$ for $\sigma \in Z_{-1,1} = \{-1, 0, 1\}$ 
satisfy the following multiplication table:

\beq
\begin{array}{l|ccc}
& P^{(+)}_\alpha & P^{(-)}_\alpha & P^{(0)}_\alpha\\
\hline
P^{(+)}_\alpha & P^{(+)}_\alpha & 0 & 0\\
P^{(-)}_\alpha & 0 & P^{(-)}_\alpha & 0\\
P^{(0)}_\alpha & 0 & 0 & P^{(0)}_\alpha\\
\end{array}
\;.
\eeq
Thus, the $P^{(\sigma)}_\alpha$ are orthogonal projection operators.

It is easy to show using the definitions of $P^{(\sigma)}_\alpha$
and $\lambda_\sigma$ for $\sigma\in Z_{-1, 1}$ that

\beq
\alpab{K} = \sum_{\sigma\in Z_{-1,1}} \lambda_\sigma P^{(\sigma)}_\alpha
\;.
\eeq
Thus,
\beq
\ln\alpab{K} = \sum_{\sigma\in Z_{-1,1}} \ln(\lambda_\sigma) P^{(\sigma)}_\alpha
\;.
\label{eq:log-kab}\eeq
Technically, we  should also add a term 
 $\ln(0) diag(0^{\#\dimv+1}, 1^{\#3 - \dimv})$
 to the right hand side of the last equation
 to account for the $3 - \dimv$ zero eigenvalues of $\alpab{K}$.
 However, we can safely ignore this infinite summand
 if we only use $\ln \alpab{K}$ in expressions
 where it is
 multiplied times $\rho_\rvab$.
The infinite summand is annihilated when $\ln \alpab{K}$ is multiplied times $\rho_\rvab$.

\subsection{$\alpab{R}$ Calculations} \label{sec:r}

To find $\alpab{R}$, we need to calculate the partial traces of $\alpab{K}$.
One gets
\beq	
\alpb{K}= \tr_\rva \alpab{K} = 
\frac{1}{N_\rvalp}\left(\frac{1}{2} + \vecalp{n}\cdot \vecsigb \right)
\;,
\eeq
and

\beq	
\alpa{K}= \tr_\rvb \alpab{K} = 
\frac{1}{N_\rvalp}\left(\frac{1}{2} + (\invtwo\vecalp{n})\cdot \vecsiga  \right)
\;,
\eeq
where 

\beq
\vecalp{n} = q \vecalp{v}
\;.
\eeq
Therefore,

\beq
\alpab{R} = \frac{\alpa{K}\alpb{K}}{w_\alpha} = 
\frac{1}{N_\rvalp}
\left(\frac{1}{2} + \vecalp{n}\cdot \vecsigb\right)
\left(\frac{1}{2} + (\invtwo\vecalp{n})\cdot \vecsiga\right)
\;.
\eeq

We also need to know $\ln \alpab{R}$. 
Using Appendix \ref{app:eigensystem}, one finds

\beqa
\ln \alpab{R} &=& 
-\ln N_\rvalp 
+ \ln\left(\frac{1}{2} + \vecalp{n}\cdot \vecsigb\right)
+ \ln\left(\frac{1}{2} + (\invtwo\vecalp{n})\cdot \vecsiga\right)
\nonumber\\
&=&
-\ln N_\rvalp  + \sum_{\xi\in Bool}\ln\left (\frac{1}{2} + (-1)^{\xi} \yy\right) {\cal P}_\xi
\;,
\eeqa
where

\beq
{\cal P}_\xi = 
\pi( \ket{\xi_{\vecalp{n}}}_\rvb ) 
+
\pi( \ket{\xi_{\invtwo \vecalp{n}}}_\rva ) 
\;
\eeq
for $\xi\in Bool$.

At this point we have calculated $\ln\alpab{R}$, but 
we have not yet 
expressed it in the desired form,
as a matrix in the Bell representation.
To do this,
we need to find the matrix elements in the Bell basis
of  the projectors
$\pi(\ket{\xi_{\vec{r}}}_\rvb)$
and $\pi(\ket{\xi_{\vec{r}}}_\rvb)$ for 
$\xi\in Bool$. These matrix elements can be found using
the techniques discussed in Section \ref{sec:bell-basis}.
One finds:  

\beqa
\pi(\ket{\xi_{\vec{r}}}_\rvb)&=& 
\frac{1}{2}
\left[1 + (-1)^\xi \vecsigb\cdot \hat{r}\right]\nonumber\\
&=& \frac{1}{2} 
+ \frac{(-1)^\xi }{2}
\left[
\begin{array}{cc}
0 & i \hat{r}^T \\
-i\hat{r} & i(\hat{r}\times \cdot)
\end{array}
\right]
\;,
\eeqa
and

\beqa
\pi(\ket{\xi_{\vec{r}}}_\rva)&=& 
\frac{1}{2}
\left[1 + (-1)^\xi \vecsiga\cdot \hat{r}\right]\nonumber\\
&=& \frac{1}{2} 
+  \frac{(-1)^\xi}{2}
\left[
\begin{array}{cc}
0 & i (\invtwo\hat{r})^T \\
-i\invtwo\hat{r} & -i(\invtwo\hat{r}\times\cdot)
\end{array}
\right]
\;.
\eeqa
Putting all this together, we get

\beq
\ln \alpab{R} = -\ln N_\rvalp 
+ \ln\left( (\frac{1}{2} + \yy)(\frac{1}{2} - \yy)\right)
+ \ln\left( \frac{\frac{1}{2} + \yy}{\frac{1}{2} - \yy}\right)
\left[
\begin{array}{cc}
0 & i \hat{ n}^{\alpha T} \\
-i\hat{ n}^{\alpha} & 0
\end{array}
\right]
\;.
\label{eq:log-rab}\eeq

\subsection{$E_{pure}$ and $E_{mixed}$ Calculations}\label{sec:entan}

Recall from Ref.\cite{Tuc} that the following Lagrangian ${\cal L}$ 
must be minimized to obtain both {$E_{pure}$ and $E_{mixed}$:
\beq
{\cal L} = l_K - l_R
\;,
\eeq
where

\beq
l_K = \sum_\alpha \tr_\rvab (\alpab{K}\ln \alpab{K})
\;,
\eeq
and

\beq
l_R = \sum_\alpha \tr_\rvab (\alpab{K}\ln \alpab{R})
\;.
\eeq
Using the results of previous sections, one finds

\beq
l_K =
-\ln N_\rvalp + \sum_{\sigma = \pm}
\lambda'_\sigma \ln (\lambda'_\sigma) + (\dimv - 1) \lambda'_0 \ln (\lambda'_0)
\;,
\eeq
where $\lambda'_\sigma = N_\rvalp \lambda_\sigma$ for $\sigma\in Z_{-1,1}$, 
and the $\lambda_\sigma$
are just the  eigenvalues of $\alpab{K}$ that we found earlier.
One also finds that

\beq
l_R =
-\ln N_\rvalp - 2h_e(\frac{1}{2} + \yy)
\;.
\eeq
Putting all this together, we get

\beq
{\cal L} =
\left\{
\begin{array}{l}
\sum_{\sigma=\pm}(\frac{u}{2} + \sigma \xx) \ln( \frac{u}{2} + \sigma \xx)\\
+(\dimv - 1) \epsilon m_1 \ln (\epsilon m_1)\\ 
+ 2 h_e(\frac{1}{2} + \yy)
\end{array}
\right.
\;.
\label{eq:gen-lagrangian}\eeq
Next we will use Eq.(\ref{eq:gen-lagrangian}) 
to calculate entanglement $E = min ({\cal L})/(2 \ln(2))$
for pure and mixed minimizations. 

\noindent {\bf (case 1)Pure Min.}

In the case of pure minimization, one has
$q=\sqrt{m_0 m_1}$ and $\epsilon=0$. Thus, the auxiliary 
parameters defined by Eqs.(\ref{eq:aux-params}) reduce to:
$\eta = \dimv$,
$u=1$,
$k = (m_0 - \dimv m_1)/2$,
$\yy = \sqrt{m_0 (\dimv m_1)} = \sqrt{m_0(1-m_0)}$,
and $\xx = 1/2$. Thus, from Eq.(\ref{eq:gen-lagrangian}), we get

\beq
E_{pure} = h\left(
\frac{1}{2} + \yy
\right)
\;.
\label{eq:e-pure}\eeq
If we define the {\it concurrence} $C$ for 
this case to be:

\beq
C = |2m_0 -1|
\;,
\eeq
then Eq.(\ref{eq:e-pure}) can be rewritten as
in Ref.\cite{Ben}:

\beq
E_{pure} = h\left(
\frac{1 + \sqrt{1-C^2}}{2} 
\right)
\;.
\eeq

\noindent {\bf (case 2)Mixed Min.}

\begin{figure}[h]
	\begin{center}
	\epsfig{file=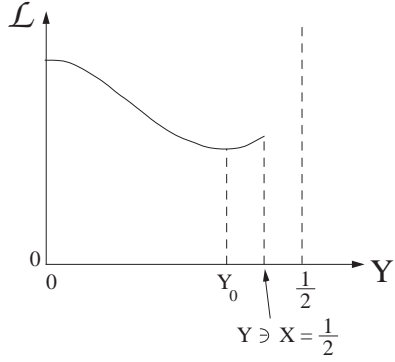, height=2.0in}
	\caption{Plot of ${\cal L}$ versus 
	$\yy$ at fixed $k$ when  
	$\dimv=1$. See Eq.(\ref{eq:L-mixed-dimv1}).}
	\label{fig:L-vs-y}
	\end{center}
\end{figure}

For mixed minimization, the constraints $q = \sqrt{m_0 m_1}$ and $\epsilon = 0$ 
are no longer required in order
to make $\alpab{K}$ separable. We can choose
$q$ and $\epsilon$ so as to minimize ${\cal L}$ given by Eq.(\ref{eq:gen-lagrangian}).
Treating ${\cal L}$ as a function of $q$ and $\epsilon$
and setting its partials to zero, we get
the following two constraints

\begin{subequations}
\beq
\pder{{\cal L}}{\epsilon} = 0 
\;\Rightarrow\; 
\left \{
\begin{array}{l}
{\rm either}\;\;\;
\ln (\epsilon m_1) =
\sum_{\sigma = \pm}
(\frac{1}{2} - \frac{\sigma k}{2 \xx} ) \ln (\frac{u}{2} + \sigma \xx)\\
{\rm or}\;\;\;
m_1(\dimv-1)=0
\end{array}
\right.
\;,
\label{eq:L-partial-eps}\eeq
and

\beq
\pder{{\cal L}}{q} = 0 
\;\Rightarrow\;
\frac{1}{2\xx} \ln
\left(
\frac{ \frac{u}{2} - \xx }{ \frac{u}{2} + \xx }
\right)
=
\frac{1}{\yy} \ln
\left(
\frac{ \frac{1}{2} - \yy }{ \frac{1}{2} + \yy }
\right)
\;.
\label{eq:L-partial-q}\eeq
\label{eq:L-partials}\end{subequations}
In general,

\beq
E_{mixed} = \frac{{\cal L}}{2 \ln 2}
\;,
\eeq
with ${\cal L}$ given by Eq.(\ref{eq:gen-lagrangian}), 
subject to the constraints Eqs.(\ref{eq:L-partials}). 

Eqs.(\ref{eq:L-partials}) with $\dimv\neq 1$ constitute a system of two nonlinear equations 
which one would like to solve for the two unknowns $\epsilon$ and $q$.
As far as we know, this  ``$\epsilon-q$ System" cannot be solved exactly in 
closed form-- its roots can only be found using numerical techniques. 
See Appendix \ref{app:e-q-sys} if interested in
an approximate analytical solution of the system.

The case of mixed min simplifies considerably
when $\dimv=1$. In that case $\alpab{K}$
and ${\cal L}$ are
independent of $\epsilon$, so we need only minimize over $q$.
${\cal L}$ reduces to

\beq
{\cal L} = - 
h_e\left(
\frac{1}{2} + \sqrt{k^2 + \yy^2}
\right) 
+ 2 h_e\left(
\frac{1}{2} + \yy
\right)
\;.
\label{eq:L-mixed-dimv1}\eeq
Fig.\ref{fig:L-vs-y} is a plot of ${\cal L}$ versus 
$\yy$ at fixed k, according to Eq.(\ref{eq:L-mixed-dimv1}).
The largest possible $\yy$ value corresponds to pure
minimization. ${\cal L}$ does not achieve its 
minimum at that endpoint, but rather at 
a smaller value of $\yy$, which we call
$\yy_0$ in Fig.\ref{fig:L-vs-y}.
$\yy_0$ can be determined by solving Eq.(\ref{eq:L-partial-q})
for $\yy$ as a function of $k$. We conclude that for $\dimv=1$,

\beq
E_{mixed} = 
-(\frac{1}{2}) h(\frac{1}{2} + \sqrt{k^2 + \yy_0^2})
+ h(\frac{1}{2} + \yy_0)
\;.
\eeq

\subsection{$\Delta_\rvab$ Calculations}\label{sec:delta}

Next, let us calculate $\Delta_\rvab$ 
for the cases of mixed and pure minimizations.

\noindent {\bf (case 1)Pure Min.}

The entanglement operator for pure min satisfies:

\beq
\Delta_\rvab^{pure}\ket{\psi_\alpha} =
(\ln \alpab{K} - \ln \alpab{R})\ket{\psi_\alpha}
\;,
\label{eq:def-delta-pure}\eeq
where

\beq
\ket{\psi_\alpha} =
\left(
\begin{array}{c}
\sqrt{m_0} \\
-i \sqrt{m_1} \vecalp{v}
\end{array}
\right)
\;.
\label{eq:psi-def}\eeq
Note that $\av{\psi_\alpha|\psi_\alpha} =1$.
Using the results of previous sections, one can express
the right hand side of Eq.(\ref{eq:def-delta-pure}) as follows.
Define auxiliary quantities $f$, $M$ and $a$ by

\beq
f = \sqrt{\frac{1-m_0}{m_0}}
\;,
\eeq

\beq
M = 
-\ln\left( \frac{\frac{1}{2} +  m_0 f}{\frac{1}{2} -  m_0 f}\right) 
diag\left[
f, 
\left(\frac{1}{f}\right)^{\#\dimv},
0^{\#3-\dimv}
\right]
\;,
\eeq
and

\beq
a = 
-\ln\left( (\frac{1}{2} + m_0f)(\frac{1}{2} - m_0f)\right)
\;.
\eeq
Then

\beq
\ln \alpab{R} \ket{\psi_\alpha} =
(- \ln N_\rvalp 
- M -a) \ket{\psi_\alpha}
\;.
\eeq
Since

\beq
\ln \alpab{K} \ket{\psi_\alpha} = -\ln(N_\rvalp) \ket{\psi_\alpha}
\;,
\eeq
we get

\beq
\Delta^{pure}_\rvab = M + a
\;.
\label{eq:fin-delta-pure}\eeq
Using this value for $\Delta_\rvab^{pure}$ 
and  the value for $E_{pure}$ 
that we obtained in Section \ref{sec:entan},
and also using the constraints $q=\sqrt{m_0 m_1}$,
$\epsilon=0$,
one can check that

\beq
(2 \ln 2) E_{pure} = \tr(\rho_\rvab \Delta_\rvab^{pure})
\;.
\eeq

\noindent{\bf (case 2)Mixed Min.}

The entanglement operator for mixed min satisfies:

\beq
\Delta^{mixed}_\rvab = \ln \alpab{K} - \ln \alpab{R}
\;.
\label{eq:def-delta-mixed}\eeq
Using the results of previous sections, one can express
the right hand side of Eq.(\ref{eq:def-delta-mixed}) as follows.
Define the auxiliary quantities $M$ and $a$ by

\beq
M = diag\left[\sum_{\sigma = \pm} 
\left(\frac{1}{2} + \frac{\sigma k}{2 \xx}\right)
\ln \left(\frac{u}{2} + \sigma \xx \right), 
(\ln (\epsilon m_1))^{\#\dimv}, 0^{\#3-\dimv}\right]
\;,
\eeq

\beq
a = -\ln\left( (\frac{1}{2} + \yy)(\frac{1}{2} - \yy) \right)
\;.
\eeq
Then

\beq
\Delta_\rvab^{mixed} = 
M + a
\;.
\label{eq:fin-delta-mixed}\eeq
Using this value for $\Delta_\rvab^{mixed}$, 
and the value for $E_{mixed}$ 
that we obtained in Section \ref{sec:entan},
and also using the constraints Eqs.(\ref{eq:L-partials}),
one can check that

\beq
(2 \ln 2) E_{mixed} = \tr(\rho_\rvab \Delta_\rvab^{mixed})
\;.
\eeq

\section{Implications} \label{sec:swept}

If $\Omega$ is an orthogonal matrix and we replace our
vectors $\vecalp{v}$ by $\Omega\vecalp{v}$, then $\alpab{K}$
changes but the value of the entanglement doesn't. Thus, 
there is a continuum of
possible $\alpab{K}$'s that minimizes ${\cal L}(K, K)$.
It is  convenient to define a {\it pure-min-entanglement orbit}
(ditto, {\it mixed-min-entanglement orbit})
as a set of all $\alpab{K}\in {\cal K}_{pure}$ 
(ditto, $\alpab{K}\in {\cal K}_{mixed}$) 
which are stationary points of ${\cal L}(K, K)$, and 
which give the same value for ${\cal L}(K, K)$.
Only one pure-min (ditto, mixed-min) orbit is a global minimum of ${\cal L}(K, K)$.
Any $\alpab{K}$ (or any orbit) 
whose $\Delta_\rvab$ is independent of $\alpha$, 
will be said to be {\it $\alpha$-insensitive}
(or just insensitive for short).
 
In previous sections, we found a pure-min insensitive $\alpab{K}$ 
whose concurrence is $|2m_0 -1|$,
regardless of whether $m_0>\frac{1}{2}$ or not.
On the other hand, in Ref.\cite{Ben}, Bennett et al found
a different $\alpab{K}$ whose concurrence is zero
when $m_0\leq \frac{1}{2}$.
(In fact, Ref.\cite{Ben}
 shows that {\it any} Bell Mixture,
 not just the Werner states that we are considering here,
 must have zero entanglement of formation
when the largest weight $m_0\leq \frac{1}{2}$).
Thus, for the Werner states that we are considering here,
there exist at least two pure-min insensitive orbits
when $m_0< \frac{1}{2}$.

In Ref.\cite{Tuc}, we claimed that
there is only one orbit, the global minimum of ${\cal L}(K, K)$, 
that is insensitive.
 This paper has given a counterexample to that claim.
It appears from this paper that the insensitivity condition
is necessary for the global minimum orbit, but it is not always sufficient.
This leads one to wonder why and when the sufficiency part of the proof in Ref.\cite{Tuc}
breaks down. The breakdown may be due to the fact that
the proof treats in a cavalier manner 
the infinities produced by taking the log of zero eigenvalues.

If we generalize our ansatz for $\alpab{K}$  so that the 
$\vecalp{v}$'s can be complex, can we find any more insensitive orbits?
We try to answer this question in Appendices \ref{app:stat-points-of-c}
and \ref{app:v-complex}.
We find that going from 
real to complex $\vecalp{v}$'s yields
new insensitive orbits in the pure min
but not in the mixed min cases.
We also find that for the $\vecalp{v}$-complex ansatz, 
there exist a countable number of pure-min insensitive orbits, 
and each of these corresponds to a stationary point of the pre-concurrence.

\begin{appendix}
\section{Appendix: Eigensystem of 2 Dimensional\\Hermitian matrix}\label{app:eigensystem}

Consider any 2 by 2 Hermitian matrix $\tilde{n}$.
One can always express it as $\tilde{n} = n_0 + \vec{n}\cdot \vec{\sigma}$,
where $n_0$ and $\vec{n}$ are real. 
The eigensystem of $\tilde{n}$ follows immediately from the following easily proven identity:

\beq
n_0 + \vec{n}\cdot \vec{\sigma} =
(n_0 + \sqrt{\vec{n}^2}) 
\left(
\frac{1 + \hat{n}\cdot \vec{\sigma}}{2}
\right)
+
(n_0 - \sqrt{\vec{n}^2}) 
\left(
\frac{1 - \hat{n}\cdot \vec{\sigma}}{2}
\right)
\;,
\label{eq:diag-two-by_two}\eeq
where $\hat{n} = \vec{n}/\sqrt{\vec{n}^2}$.
Define
\beq
P_\pm = \frac{1 \pm \hat{n}\cdot \vec{\sigma}}{2}
\;.
\label{eq:z-projs}\eeq
Then 

\beq
P_+ P_- = P_- P_+ = 0
\;,
\eeq
and

\beq
(P_\pm)^2 = P_\pm
\;.
\eeq
Thus,  $P_+$ and $P_-$ are the projectors onto the two eigenspaces of $\tilde{n}$
with respective eigenvalues $n_0 + \sqrt{\vec{n}^2}$ and $n_0 - \sqrt{\vec{n}^2}$.

An alternative,  more tedious way of finding the eigensystem of $\tilde{n}$
is to rotate the equations $\sigma_z\ket{0}=\ket{0}$ and $\sigma_z\ket{1}=-\ket{1}$.
Define a rotation vector $\vec{\theta}$ by:
\beq
\vec{\theta} = \theta \frac{\hat{z}\times \vec{n}}{|\hat{z}\times \vec{n}|}
\;,\;\;
\theta = {\arccos} \frac{n_3}{|\vec{n}|}
\;,
\eeq
The spin up and down states along the $\vec{n}$ direction can be obtained
in terms of those along the $\hat{z}$ by:

\beq
\ket{0_{\vec{n}}} = 
\exp( -i \frac{\vec{\sigma}\cdot \vec{\theta}}{2} ) \ket{0}
\;,
\eeq
and

\beq
\ket{1_{\vec{n}}} = 
\exp( -i \frac{\vec{\sigma}\cdot \vec{\theta}}{2} ) \ket{1}
\;.
\eeq
One can show that the projectors $P_\pm$
defined by Eq.(\ref{eq:z-projs}) satisfy:

\beq
P_+ = \proj{0_{\vec{n}}}
\;,
\eeq
and

\beq
P_- = \proj{1_{\vec{n}}}
\;.
\eeq

\section{Appendix: Approximate Solution of $\epsilon-q$ System}\label{app:e-q-sys}

In this appendix, we will assume $\dimv>1$.
Eqs.(\ref{eq:L-partials}) constitute a system of two nonlinear equations 
which one would like to solve for the two unknowns $\epsilon$ and $q$.
As far as we know, this  ``$\epsilon-q$ System" cannot be solved exactly in 
closed form-- its roots can only be found using numerical techniques. 
Nevertheless, as we will show next, it is possible to 
get approximate analytical expressions for its roots.

It is easy to show that the $\epsilon-q$ System is solved
exactly by $\mathbf (\epsilon, q) = (1, 0)$. This root, however,
does not yield the global minimum mixed-min orbit, so it is of
little interest to us. It can be rejected if we restrict
our attention to roots for which the regulator $q$ is nonzero.

From the numerical results obtained with Causa Com\'{u}n 
and discussed in Ref.\cite{Tuc}, we expect that the $\epsilon-q$ System
has a second root which is very close to the pure min case,
for which $\mathbf (\epsilon, q) = (0, \sqrt{m_0 m_1})$.
The rest of this appendix will be devoted to finding 
an approximate value for this second root.
Taylor expansion at $(\epsilon, q) = (0, \sqrt{m_0 m_1})$
is not possible since both equations of the $\epsilon-q$ System
are non-analytic at that point.
Another type of approximation is called for.

We begin by rewriting the $\epsilon-q$ System
in the following equivalent form:

\begin{subequations}
\beq
\epsilon m_1 =
\left(\frac{u}{2} + \xx\right)^{\frac{1}{2} - \frac{k}{2\xx}}
\left(\frac{u}{2} - \xx\right)^{\frac{1}{2} + \frac{k}{2\xx}}
\;,
\eeq

\beq
\left(
\frac{ \frac{u}{2} - \xx }{ \frac{u}{2} + \xx }
\right)^\yy
=
\left(
\frac{ \frac{1}{2} - \yy }{ \frac{1}{2} + \yy }
\right)^{2\xx}
\;.
\eeq
\label{eq:e-q-sys-powers}\end{subequations}
When 
$\epsilon\approx 0$ and 
$q\approx \sqrt{m_0 m_1}$, one has that
$u\approx u^0 \equiv 1$,
$k\approx k^0  \equiv m_0 -\frac{1}{2}$,
$\yy\approx \yy^0  \equiv \sqrt{m_0 (1-m_0)}$,
and
$\xx\approx \xx^0  \equiv \frac{1}{2}$. 
Our approximation consists of replacing 
Eqs.(\ref{eq:e-q-sys-powers}) by the following two equations:

\begin{subequations}
\beq
\epsilon m_1 =
\left(\frac{u^0}{2} +  \xx^0\right)^{\frac{1}{2} - \frac{k^0}{2\xx^0}}
\left(\frac{u}{2} - \xx\right)^{\frac{1}{2} + \frac{k^0}{2\xx^0}}
\;,
\eeq

\beq
\left(
\frac{ \frac{u}{2} - \xx }{ \frac{u^0}{2} + \xx^0 }
\right)^{\yy^0}
=
\left(
\frac{ \frac{1}{2} - \yy }{ \frac{1}{2} + \yy^0 }
\right)^{2\xx^0}
\;.
\eeq
\label{eq:e-q-sys-approx}\end{subequations}
In effect, what we are doing is setting $\epsilon=0$ and
$q= \sqrt{m_0 m_1}$ everywhere in Eqs.(\ref{eq:e-q-sys-powers}) except where
a very small number (namely, 
$\frac{u}{2} - \xx$ and $\frac{1}{2} - \yy$) appears raised to a power.
Since $\frac{1}{2}-Y^0=0$ iff $m_0= \frac{1}{2}$, this approximation is expected
to work best in the vicinity of $m_0= \frac{1}{2}$.
Define the quantity $\rho$ by:

\beq
\rho = m_0 m_1 \eta - q^2 \dimv = (\frac{u}{2})^2 -\xx^2
\;.
\eeq
Eqs.(\ref{eq:e-q-sys-approx}) can be expressed in terms of $\rho$ as follows:

\begin{subequations}
\beq
\epsilon m_1 = \rho^{m_0}
\;,
\eeq

\beq
\rho^{\yy^0} (\frac{1}{2} + \yy^0) = \frac{1}{2} - q \sqrt{\dimv}
\;.
\eeq
\label{eq:e-q-sys-rho}\end{subequations}
Eqs.(\ref{eq:e-q-sys-rho}) imply

\begin{subequations}
\beq
m_0 m_1 \eta 
= m_0 m_1 \left[ \dimv - \epsilon(\dimv - 1)\right]
= m_0 m_1 \left[ \dimv - \frac{\rho^{m_0}}{m_1}(\dimv - 1)\right]
\;,
\eeq
and

\beq
q^2\dimv = \left[ \frac{1}{2} - \rho^{\yy^0}(\frac{1}{2} + \yy^0)\right]^2
\;.
\eeq
\label{eq:rho-terms}\end{subequations}
Now the left hand sides of Eqs.(\ref{eq:rho-terms}) are just the two terms 
whose difference defines $\rho$. Therefore,

\beq
\rho = 
(\yy^0)^2 \left[ 1 - \frac{\rho^{m_0}}{m_1}(1 - \frac{1}{\dimv})\right] 
- \left[ \frac{1}{2} - \rho^{\yy^0}(\frac{1}{2} + \yy^0)\right]^2
\;.
\eeq
Motivated by the last equation, we define a function $f$ of $\rho$ by: 

\beq
f(\rho) = -\rho
+ (\yy^0)^2 \left[ 1 - \frac{\rho^{m_0}}{1-m_0}(\dimv - 1)\right] 
- \left[ \frac{1}{2} - \rho^{\yy^0}(\frac{1}{2} + \yy^0)\right]^2
\;.
\eeq
Eqs.(\ref{eq:e-q-sys-rho}) can rewritten as follows, so that they express $\epsilon$ and $q$
in terms of $\rho$:

\begin{subequations}
\beq
\epsilon = \frac{\rho^{m_0}}{1-m_0}\dimv
\;,
\eeq

\beq
q = \frac{1}{\sqrt{\dimv}}\left[ \frac{1}{2} - \rho^{\yy^0}(\frac{1}{2} + \yy^0)\right]
\;.
\eeq
\label{eq:e-q-as-func-rho}\end{subequations}
Clearly, given any $\rho_{root}$ for which $f(\rho_{root})=0$ , one can use Eqs.(\ref{eq:e-q-as-func-rho})
to calculate a point $(\epsilon, q)_{root}$ that approximately 
satisfies the original $\epsilon-q$ System.
If  $(\epsilon, q)_{root}$ yields values for $u$, $k$, $\xx$ and $\yy$
that are close to  $u^0$, $k^0$, $\xx^0$ and $\yy^0$, respectively,
then our original assumptions are vindicated and we say the approximation is self consistent.

\begin{figure}[h]
	\begin{center}
	\epsfig{file=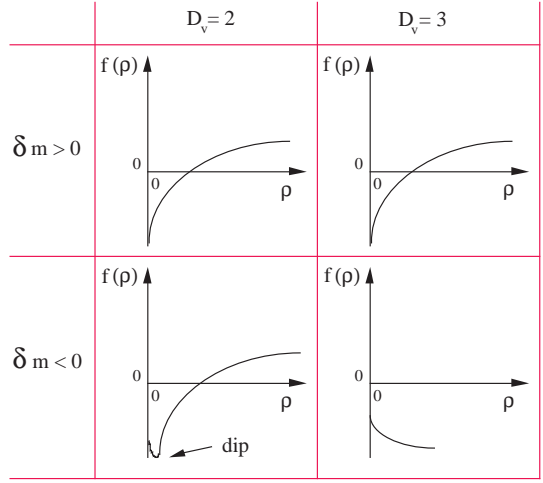, height=2.5in}
	\caption{Schematic plots of $f(\rho)$. Four cases depicted depending on 
	whether $\dimv$ equals 2 or 3 and whether $\delta m$ is greater or smaller than zero. }
	\label{fig:f-rho}
	\end{center}
\end{figure}

Let $m_0 = \frac{1}{2} + \delta m$.
We wrote a simple Excel spreadsheet in which we plotted $f(\rho)$ as a function of $\rho$, 
with inputs $m_0$ and $\dimv$. 
Fig.\ref{fig:f-rho} shows schematically what we found. 
For $\dimv=2$, $f(\rho)$ has a zero regardless of the sign of $\delta m $.
For $\dimv=3$, $f(\rho)$ has a zero for $\delta m >0$ but not for  $\delta m <0$.

We also calculated analytical approximations for 
$f$, and for its first and second derivatives, in the limit
of small $\rho$ and small $|\delta m|$: 

\beq
f(0) = (\yy^0)^2 - \frac{1}{4}\leq 0
\;,
\eeq

\beq
\frac{\partial f}{\partial\rho} 
\rarrow \frac{1}{\sqrt{\rho}} 
\left[ \frac{-1}{4} (\dimv -1) \rho^{\delta m} + \frac{1}{2} \right]
\rarrow
\left\{
\begin{array}{l}
\frac{1}{2\sqrt{\rho}}\;\;{\rm if}\;\; \delta m > 0 \\
\\
\frac{-(\dimv-1)}{4\rho^{\frac{1}{2} + |\delta m|}}\;\;{\rm if}\;\; \delta m < 0 
\end{array}
\right.
\;,
\eeq

\beq
\frac{\partial^2 f}{\partial\rho^2} 
\rarrow \frac{1}{\rho^{\frac{3}{2}}} 
\left[ \frac{-1}{4} (\dimv -1)(\delta m - \frac{1}{2}) \rho^{\delta m} - \frac{1}{4} \right]
\rarrow
\left\{
\begin{array}{l}
\frac{-1}{4\rho^{\frac{3}{2}}}\;\;{\rm if}\;\; \delta m > 0 \\
\\
\frac{(\dimv-1)}{8\rho^{\frac{3}{2} + |\delta m|}}\;\;{\rm if}\;\; \delta m < 0 
\end{array}
\right.
\;.
\eeq

\section{Appendix: Stationary Points of \\the Pre-concurrence}\label{app:stat-points-of-c}

Define the {\it pre-concurrence amplitude} $\gamma_\theta$ by 

\beq
\gamma_\theta = \sum_{j=0}^n e^{i\theta_j} m_j
\;,
\eeq
where $m_j\geq 0$ for all $j\in Z_{0,n}$, $\sum_{j=0}^n m_j = 1$, the $\theta_j$
are real numbers, and we fix $\theta_0 = 0$. Next define 
the {\it pre-concurrence} $C_\theta$ by

\beq
C_\theta = |\gamma_\theta|
\;.
\eeq
The global minimum of $C_\theta$ (at $n=3$) over all phases $\vec{\theta}$
 is called the {\it concurrence}.
This global minimum arose in our calculation 
of the pure min entanglement of a two qubit system.
One wonders if $C_\theta$ has other stationary points, and
whether they play a role in our theory.
In this appendix, we will find the stationary points of $C_\theta$.
In Appendix \ref{app:v-complex}, we will show that such stationary 
points are indeed very relevant to our theory.
 
\begin{figure}[h]
	\begin{center}
	\epsfig{file=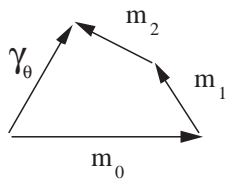, height=1in}
	\caption{$\gamma_\theta$ is a ``fractured unit vector".}
	\label{fig:frac-unit-vec}
	\end{center}
\end{figure}

\begin{figure}[h]
	\begin{center}
	\epsfig{file=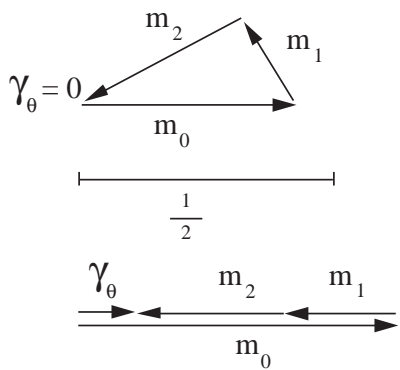, height=2.0in}
	\caption{$\min |\gamma_\theta|$ equals zero if $m_0 \leq \frac{1}{2}$
	and $m_0 - \sum_{j=1}^n m_j$ if $m_0 >\frac{1}{2}$.}
	\label{fig:min-gamma}
	\end{center}
\end{figure}

From Fig.\ref{fig:frac-unit-vec}, we see that $\gamma$ is a ``fractured unit vector" in the complex plane:
it equals the vector sum of segments whose lengths $m_j$ add to one.
From this geometrical picture, the smallest possible 
length for $\gamma_\theta$ is clear.
 As shown in Fig.\ref{fig:min-gamma}, if $m_0\leq \frac{1}{2}$,
then $\min |\gamma_\theta| = 0$. If $m_0 > \frac{1}{2}$, 
then $\min |\gamma_\theta| = m_0 - \sum_{j=1}^n m_j$.
Are there any other stationary points of $C_\theta$? Yes.

Any stationary point $\vec{\theta}$ of $C_\theta$ must satisfy:
\beq
\delta C_\theta^2 = \delta(\gamma_\theta\gamma_\theta^*)= 
i\sum_{j=1}^n \delta\theta_j m_j (e^{i\theta_j} \gamma_\theta^* - e^{-i\theta_j}\gamma_\theta) = 0
\;.
\eeq
Hence,

\beq
m_j|\gamma_\theta|\sin\left(\theta_j - \angle(\gamma_\theta)\right)=0
\;,
\eeq
for all $j\in Z_{1,n}$. The last equation is satisfied iff either
$\gamma_\theta=0$, or,  for all $j\in Z_{1,n}$ 
such that $m_j\neq 0$, $\theta_j = \angle(\gamma_\theta) + \pi b_j + 2\pi n_j $, 
where $b_j\in Bool$ and the $n_j$ are integers. Hence, the set of stationary values which
$C_\theta$ can assume is given by 

\beq
\left\{ |\sum_{j=0}^n (-1)^{b_j} m_j| : b_j\in Bool, b_0 = 0\right\}
\;,
\label{eq:conc-stat-pts}\eeq
together with $C_\theta=0$, which, however, is only possible when $m_0\leq \frac{1}{2}$.

\begin{figure}[h]
	\begin{center}
	\epsfig{file=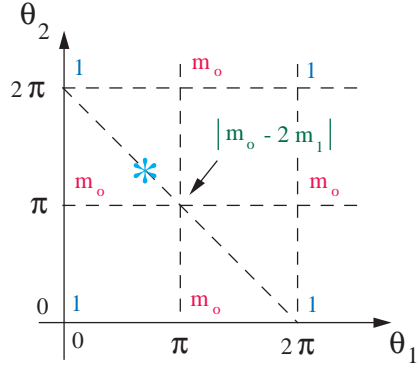, height=2.0in}
	\caption{Values of $C_\theta(\theta_1, \theta_2)$ shown in color.
	$C_\theta$ is periodic in $\theta_1$ and $\theta_2$. Its minimum value is 
	$|m_0 - 2m_1|$ if $m_0 \geq \frac{1}{2}$ and zero otherwise. $C_\theta = 0$
	(possible only if $m_0 \leq \frac{1}{2}$) is indicated by an asterisk.}
	\label{fig:concurrence}
	\end{center}
\end{figure}

Fig.\ref{fig:concurrence} shows a plot of   
\beq
C_\theta(\theta_1, \theta_2) = | m_0 + m_1 (e^{i\theta_1} + e^{i\theta_2})|
\;; 
\eeq
that is, $C_\theta$ when $n=2$ and $m_1 = m_2$.

\section{Appendix: Complex $\vecalp{v}$'s}\label{app:v-complex}

Recall that an orbit is a set of all $\alpab{K}$ which give a common
stationary value for ${\cal L}(K, K)$. And if
$\Delta_\rvab$ for an orbit is independent of $\alpha$,
we say that the orbit is insensitive.
As was pointed out in Section \ref{sec:swept},
the results of this paper imply that there can be more than 
one pure-min insensitive orbit.
If we generalize our ansatz for $\alpab{K}$  so that the 
$\vecalp{v}$'s can be complex, can we find any more insensitive orbits?
We know that the global minimum of the pre-concurrence corresponds
to an insensitive orbit.
Do the other stationary points of the pre-concurrence also correspond
to insensitive orbits? 
In this appendix we will show that 
going from 
real to complex $\vecalp{v}$'s yields
new insensitive orbits in the pure min
but not in the mixed min cases.
We will also show  that for the $\vecalp{v}$-complex ansatz, 
there exist a countable number of pure-min insensitive orbits, 
and each of these corresponds to a stationary point of the pre-concurrence.

Let $U^\alpha$ be defined by
\beq
U^\alpha = diag(e^{i\phi^\alpha_1}, e^{i\phi^\alpha_2}, e^{i\phi^\alpha_3})
\;,
\eeq
where the $\phi^\alpha_j$ are real. Let

\beq
\bar{U}^\alpha = 
\left[
\begin{array}{cc}
1 & 0 \\
0 & U^\alpha
\end{array}
\right ]
\;.
\eeq
One can define complex vectors $\vecalp{\zeta}$ in terms 
the real ones $\vecalp{v}$ by:
\beq
\vecalp{\zeta} = U^\alpha  \vecalp{v}
\;.
\eeq
Note that by virtue of Eqs.(\ref{eq:v-constraints}),
$\vecalpdag{\zeta}\vecalp{\zeta} = \dimv$,
$\sum_\alpha \vecalp{\zeta} = 0$, and
$\sum_\alpha \vecalp{\zeta}\vecalpdag{\zeta} = N_\rvalp I_v$.
Our new ansatz for $\alpab{K}$ is  defined in terms of the 
old one by:
\beq
\alpab{\tilde{K}} = \bar{U}^\alpha \alpab{K} (\bar{U}^\alpha)^*
= (\alpab{K})_{\vecalp{v}\rarrow \vecalp{\zeta}}
\;.
\eeq
Henceforth, as we have just done, when we need to distinguish between a new quantity
and the corresponding old one, we will indicate the new one
by a tilde. (By old we mean with the $\vecalp{v}$ real,
and by new, with the $\vecalp{v}$ complex.) Sometimes,
if it is clear from the context that we are speaking of new 
quantities, we will drop the tildes. Alternatively, sometimes we will introduce 
a new symbol for a new quantity to distinguish it from the corresponding old one, 
as we did by introducing the symbol
$\vecalp{\zeta}$ to represent the new $\vecalp{v}$.

It is convenient to define $z_0$ and $\vecalp{z}$ by:

\beq
z_0 = \sqrt{m_0}
\;,
\eeq

\beq
\vecalp{z} = -i \sqrt{m_1} \vecalp{\zeta}
\;.
\eeq
Note that $(z_0)^2  + \vecalp{z}\cdot \vecalpstar{z} = 1$.

One can define a {\it concurrence amplitude} $\gamma$ by
\beqa
\gamma 
&=& (z_0)^2  + (\vecalp{z})^2 \nonumber\\
&=& m_0  - m_1 (\vecalp{\zeta})^2 \nonumber\\
&=& m_0  - m_1 \sum_{j=1}^3 e^{i 2\phi^\alpha_j} (v^\alpha_j)^2
\;.
\label{eq:complex-concurrence}\eeqa
The {\it concurrence} $C$ is defined as the absolute value of $\gamma$:
\beq
C = |\gamma|
\;.
\eeq
Note that in general, $\gamma$ and $C$ depend on $\alpha$, However,
as will soon become apparent, they must be independent of $\alpha$ for
any insensitive orbit.

Next we will calculate the new
$\alpab{\tilde{K}}$, $\alpab{\tilde{R}}$ and $\tilde{\Delta}_\rvab$.
The calculation is very similar to that of the old
$\alpab{K}$, $\alpab{R}$ and $\Delta_\rvab$
presented in previous sections.

\noindent{\bf (a) $\alpab{\tilde{K}}$ Calculations}

Since $\alpab{\tilde{K}}$ and $\alpab{K}$ are simply related by a unitary transformation,
it is clear from previous results for $\alpab{K}$ (see Section \ref{sec:k}) that
 
\beq
\ln\alpab{\tilde{K}} = -\ln N_\rvalp + \sum_{\sigma\in Z_{-1,1}} \ln(\lambda'_\sigma) \tilde{P}^{(\sigma)}_\alpha
\;,
\eeq
where

\beq
\tilde{P}^{(\sigma)}_\alpha = \bar{U}^\alpha P^{(\sigma)}_\alpha (\bar{U}^\alpha)^*
= (P^{(\sigma)}_\alpha)_{\vecalp{v}\rarrow \vecalp{\zeta}}
\;.
\eeq
Written more explicitly, this becomes

\beq
\ln\alpab{\tilde{K}} = 
\left\{
\begin{array}{l}
-\ln N_\rvalp\\
+ \sum_{\sigma=\pm} \ln( \frac{u}{2} + \sigma \xx)
\left(
\begin{array}{cc}
\frac{1}{2} + \frac{\sigma k}{2\xx} & \frac{\sigma i q \vecalpdag{\zeta}}{2\xx}\\
\frac{-\sigma i q \vecalp{\zeta}}{2\xx} & (\frac{1}{2} - \frac{\sigma k}{2\xx})\frac{\vecalp{\zeta}\vecalpdag{\zeta}}{\dimv}
\end{array}
\right) \\
+ \ln(\epsilon m_1)
\left(
\begin{array}{cc}
0 & 0 \\
0 & I_v - \frac{\vecalp{\zeta}\vecalpdag{\zeta}}{\dimv}
\end{array}
\right)
\end{array}
\right.
\;.
\label{eq:ln-k-tilde}\eeq

\noindent{\bf (b) $\alpab{\tilde{R}}$ Calculations} 

We proceed as we did in Section \ref{sec:r} where we calculated $\alpab{R}$. Now
we find that:

\beq	
\alpb{\tilde{K}}= \tr_\rva \alpab{\tilde{K}} = 
\frac{1}{N_\rvalp}\left(\frac{1}{2} + (\vecalp{n} + \Delta \vecalp{n}) \cdot \vecsigb \right)
\;,
\label{eq:plus-delta}\eeq
and

\beq	
\alpa{\tilde{K}}= \tr_\rvb \alpab{\tilde{K}} = 
\frac{1}{N_\rvalp}\left(\frac{1}{2} + [\invtwo(\vecalp{n} - \Delta\vecalp{n})]\cdot \vecsiga  \right)
\;,
\label{eq:minus-delta}\eeq
where

\beq
\vecalp{n} = \frac{i}{2} (\vecalp{z} - \vecalpstar{z})z_0 \frac{q}{\sqrt{m_0 m_1}}
\;,
\eeq
and

\beq
\Delta \vecalp{n} = 
\frac{i}{2} (\vecalp{z} \times \vecalpstar{z}) (1- \epsilon)
\;.
\eeq
Note that the $\Delta\vecalp{n}$ enters with opposite signs in Eqs.(\ref{eq:plus-delta}) and
(\ref{eq:minus-delta}).
Proceeding as we did in Section \ref{sec:r},
we finally find

\beq
\ln \alpab{\tilde{R}} = 
\left\{
\begin{array}{l}
-\ln N_\rvalp \\
+ \ln\left( (\frac{1}{2} + \tilde{\yy})(\frac{1}{2} - \tilde{\yy})\right)\\
+ \ln\left( \frac{\frac{1}{2} + \tilde{\yy}}{\frac{1}{2} - \tilde{\yy}}\right)
\frac{1}{\tilde{\yy}}
\left[
\begin{array}{cc}
0 & i \vec{ n}^{\alpha T} \\
-i\vec{ n}^{\alpha} & i \Delta\vecalp{n} \times\cdot
\end{array}
\right]
\end{array}
\right.
\;,
\label{eq:ln-r-tilde}\eeq
where

\beq
\tilde{\yy} = | \vecalp{n} + \Delta \vecalp{n} |
\;. 
\eeq

Note that in general, $\tilde{\yy}$ depends on $\alpha$, However,
as will soon become apparent,  $\tilde{\yy}$ must be independent of $\alpha$ for
any insensitive orbit. Note also that $\vecalp{n}\cdot \Delta \vecalp{n}=0$ 
so $\tilde{\yy}^2= |\vecalp{n}|^2 + | \Delta \vecalp{n} |^2$.
After some algebra, one can show that:
 
\beq
|\vecalp{n}|^2 = \left( \frac{q^2}{m_0 m_1} \right) \frac{m_0}{2} ( 1 - \gamma_r )
\;,
\label{eq:n-mag-sq}\eeq
and

\beqa
| \Delta \vecalp{n} |^2  
&=& (1-\epsilon)^2 \left\{ \frac{1 - |\gamma|^2}{4} -\frac{m_0}{2}(1 - \gamma_r)\right\}\nonumber\\
&=& \frac{(1-\epsilon)^2}{4} 
\left\{ (1 - \gamma_r)[\gamma_r - (2m_0-1)] - \gamma_i^2 \right\}
\;,
\label{eq:delta-n-mag-sq}\eeqa
where $\gamma_r$ and $\gamma_i$ are the real and imaginary parts of $\gamma$.
We see that when $\gamma = 1$ for all $\alpha$, $\vecalp{n}= \Delta \vecalp{n} = 0$ for
all $\alpha$. In the other extreme, when $\gamma=0$, one gets
$|\vecalp{n}|^2 = \frac{q^2}{2 m_1}$
and
$| \Delta \vecalp{n} |^2  = \frac{(1-\epsilon)^2}{4}(1-2m_0)$.
 If besides $\gamma=0$, one assumes the pure min constraints $q=\sqrt{m_0 m_1}$, $\epsilon=0$,
then $|\vecalp{n}|^2 = m_0/2$, $| \Delta \vecalp{n} |^2 = (1-2m_0)/4$, 
so $\tilde{\yy} = |\vecalp{n} + \Delta \vecalp{n}| =  \frac{1}{2}$. Recall that for the
real-$\vecalp{v}$ ansatz, one has $\gamma= |2m_0-1|$ and $\yy = \sqrt{m_0(1-m_0)}$,
and therefore, $\gamma=0$ implies $m_0 = \yy = \frac{1}{2}$. We see that in the complex-$\vecalp{v}$ ansatz,
it is possible to have $\gamma=0$, $\tilde{\yy}=\frac{1}{2}$, and $m_0 \neq \frac{1}{2}$.

\noindent{\bf (c) $\tilde{\Delta}_\rvab$ Calculations}

For the mixed min case, $\tilde{\Delta}_\rvab^{mixed}$ is simply

\beq
\tilde{\Delta}_\rvab^{mixed} = \ln \alpab{\tilde{K}} - \ln \alpab{\tilde{R}}
\;,
\label{eq:delta-mixed-tilde}\eeq
where  $\ln \alpab{\tilde{K}}$ and  $\ln \alpab{\tilde{R}}$ have just
been calculated. 

For the pure min case, $(\epsilon, q) = (0, \sqrt{m_0m_1})$ and
$\ket{\psi_\alpha}$ is defined by Eq.(\ref{eq:psi-def}), which can be expressed
in terms of $z_0$ and $\vecalp{z}$ as

\beq
\ket{\psi_\alpha} =
\left(
\begin{array}{c}
z_0 \\
\vecalp{z}
\end{array}
\right)
\;.
\eeq
One finds

\beqa
\tilde{\Delta}_\rvab^{pure}\ket{\psi_\alpha} 
&=& (\ln \alpab{\tilde{K}} - \ln \alpab{\tilde{R}})\ket{\psi_\alpha}\nonumber\\
&=&\left\{ 
\begin{array}{l}
-\ln\left( (\frac{1}{2} + \tilde{\yy})(\frac{1}{2} - \tilde{\yy})\right) \\
- \ln\left( \frac{\frac{1}{2} + \tilde{\yy}}{\frac{1}{2} - \tilde{\yy}}\right)
\frac{1}{2\tilde{\yy}}
\left(
\begin{array}{cc}
1-\gamma & 0 \\
0 & 1+\gamma (U^\alpha)^{2 *}
\end{array}
\right)
\end{array}
\right\}
\ket{\psi_\alpha}
\;.
\label{eq:delta-pure-tilde}\eeqa
As pointed out earlier in this appendix,
if $q=\sqrt{m_0 m_1}$, $\epsilon=0$, and  
$\gamma=0$, then $\tilde{\yy}= \frac{1}{2}$. In this limit,
Eq.(\ref{eq:delta-pure-tilde}) gives $\tilde{\Delta}^{pure} = 0$, as expected.

A simple Lemma: Consider an insensitive orbit for which  
$\vecalp{n}$ and $\Delta \vecalp{n}$ are independent of $\alpha$.
If this premise is satisfied, then $q=0$ and $\epsilon=1$. 
The premise is true in particular when $\gamma = 1$ for all $\alpha$. Proof:
By Eq.(\ref{eq:ln-r-tilde}) and the premise, 
$\ln \alpab{\tilde{R}}$ is independent of $\alpha$.
 Since
$\ln \alpab{\tilde{R}}$ and $\Delta_\rvab$ are independent of $\alpha$, 
$\ln \alpab{\tilde{K}}$ (and therefore $\alpab{\tilde{K}}$) 
must be independent of $\alpha$ too. From the ansatz form for 
$\alpab{\tilde{K}}$, it is clear that if $\alpab{\tilde{K}}$ is independent of $\alpha$,
then  $q=0$ and $\epsilon=1$. 
If $\gamma=1$ for all $\alpha$, then by Eqs.(\ref{eq:n-mag-sq}) and (\ref{eq:delta-n-mag-sq}),
$\vecalp{n} = \Delta \vecalp{n} = 0$. Therefore, 
$\vecalp{n}$ and $\Delta \vecalp{n}$ are independent of $\alpha$.
QED

We end this section by finding all insensitive orbits
that are possible according to the just derived formulas for
$\ln \alpab{\tilde{K}}$, $\ln \alpab{\tilde{R}}$ and $\tilde{\Delta}_\rvab$.

\noindent{\bf Pure Min Case:}

Our arguments are based on Eq.(\ref{eq:delta-pure-tilde}), which we restate here for convenience:

\beqa
\tilde{\Delta}_\rvab^{pure}\ket{\psi_\alpha} 
&=& (\ln \alpab{\tilde{K}} - \ln \alpab{\tilde{R}})\ket{\psi_\alpha}\nonumber\\
&=&\left\{ 
\begin{array}{l}
-\ln\left( (\frac{1}{2} + \tilde{\yy})(\frac{1}{2} - \tilde{\yy})\right) \\
- \ln\left( \frac{\frac{1}{2} + \tilde{\yy}}{\frac{1}{2} - \tilde{\yy}}\right)
\frac{1}{2\tilde{\yy}}
\left(
\begin{array}{cc}
1-\gamma & 0 \\
0 & 1+\gamma (U^\alpha)^{2 *}
\end{array}
\right)
\end{array}
\right\}
\ket{\psi_\alpha}
\;.
\eeqa
We will henceforth call this equation Eq.A. Note the following. 

For the right hand
side of Eq.A to be independent of $\alpha$ (for an open set of $m_0$ values), $\tilde{\yy}$ 
and $\gamma$ must be independent of $\alpha$.

If $\gamma=0$, then we must have $m_0\leq \frac{1}{2}$, as explained 
in Appendix \ref{app:stat-points-of-c}. According to that Appendix,
$\gamma=0$ is one of the possible stationary values of the pre-concurrence.

If $\gamma\neq 0$, then from Eq.A, $(U^\alpha)^2$ must be 
real and independent of $\alpha$. Thus, $e^{i2\phi^\alpha_j}=(-1)^{b_j}$
for $j\in Z_{1,3}$,
where $b_j\in Bool$ is independent of $\alpha$. This gives
$\gamma = m_0 - \sum_{j=1}^3 (-1)^{b_j} m_j$,
where $m_j = m_1 (v^\alpha_j)^2$. In the examples given by Eqs.(\ref{eq:v-alp-egs}),
$(v^\alpha_j)^2$ is independent of $\alpha$ for all $j$.

If $b_j = 1$ for all $j$, then $\gamma=1$.
By the Lemma proven previously, this implies that $q=0$ and $\epsilon=1$.
We can exclude this case because $q=0$ and $\epsilon=1$ would not give
(except for some special values of $m_0$) a $\alpab{K}$ of the form
$w_\alpha \proj{\psi^\alpha}$.

Comparing the above results with Appendix \ref{app:stat-points-of-c},
we see that there are a countable number of pure-min insensitive orbits, 
and each of these corresponds to a stationary point of the pre-concurrence.

\noindent{\bf Mixed Min Case:}

Our arguments are based on Eqs.(\ref{eq:ln-k-tilde}), Eqs.(\ref{eq:ln-r-tilde}) and
Eqs.(\ref{eq:delta-mixed-tilde}), which we restate here for convenience:

\begin{subequations}
\beq
\ln\alpab{\tilde{K}} = 
\left\{
\begin{array}{l}
-\ln N_\rvalp\\
+ \sum_{\sigma=\pm} \ln( \frac{u}{2} + \sigma \xx)
\left(
\begin{array}{cc}
\frac{1}{2} + \frac{\sigma k}{2\xx} & \frac{\sigma i q \vecalpdag{\zeta}}{2\xx}\\
\frac{-\sigma i q \vecalp{\zeta}}{2\xx} & (\frac{1}{2} - \frac{\sigma k}{2\xx})\frac{\vecalp{\zeta}\vecalpdag{\zeta}}{\dimv}
\end{array}
\right) \\
+ \ln(\epsilon m_1)
\left(
\begin{array}{cc}
0 & 0 \\
0 & I_v - \frac{\vecalp{\zeta}\vecalpdag{\zeta}}{\dimv}
\end{array}
\right)
\end{array}
\right.
\;.
\eeq

\beq
\ln \alpab{\tilde{R}} = 
\left\{
\begin{array}{l}
-\ln N_\rvalp \\
+ \ln\left( (\frac{1}{2} + \tilde{\yy})(\frac{1}{2} - \tilde{\yy})\right)\\
+ \ln\left( \frac{\frac{1}{2} + \tilde{\yy}}{\frac{1}{2} - \tilde{\yy}}\right)
\frac{1}{\tilde{\yy}}
\left[
\begin{array}{cc}
0 & i \vec{ n}^{\alpha T} \\
-i\vec{ n}^{\alpha} & i \Delta\vecalp{n} \times\cdot
\end{array}
\right]
\end{array}
\right.
\;,
\eeq

\beq
\tilde{\Delta}_\rvab^{mixed} = \ln \alpab{\tilde{K}} - \ln \alpab{\tilde{R}}
\;.
\eeq
\end{subequations}
Henceforth, we will  call these 3 equations A, B and C, respectively.

Subtracting entries (1,0), (2,0) and (3,0) of the
right hand sides of Eqs.A and B, we get:

\beq
\ln\left( \frac{\frac{u}{2} - \xx}{\frac{u}{2} + \xx}\right)
\frac{i q \vecalp{\zeta}}{2\xx}
- \ln\left( \frac{\frac{1}{2} + \tilde{\yy}}{\frac{1}{2} - \tilde{\yy}}\right)
\frac{1}{\tilde{\yy}} (-i \vecalp{n})
\;.
\eeq
This vector must be  independent of $\alpha$. 
Since $\vecalp{n} = \frac{q}{2}(\vecalp{\zeta} +\vecalpstar{\zeta})$,
we must have $\lambda_1 \vecalp{\zeta}  + \lambda_2 \vecalpstar{\zeta} = \vec{v}$,
where $\lambda_1$, $\lambda_2$ are real numbers, and 
$\lambda_1$, $\lambda_2$, $\vec{v}$ are independent of $\alpha$.
But then $\sum_\alpha(\lambda_1 \vecalp{\zeta}  + \lambda_2 \vecalpstar{\zeta}) = 0 = \vec{v}$,
so $\lambda_1 \vecalp{\zeta}  + \lambda_2 \vecalpstar{\zeta} = 0$.
In other words, $(\lambda_1 + \lambda_2 e^{-i2\phi^\alpha_j})v^\alpha_j = 0$
 for all $\alpha$ and $j$. Assume $\lambda_1$ and $\lambda_2$
 are non-zero for an open set of $m_0$ values. Also assume
 that $v^\alpha_j \neq 0$ for all $\alpha$ and $j\in Z_{1, \dimv}$ (this is true  
 for the examples given in Eqs.(\ref{eq:v-alp-egs}). Then 
 $e^{-i2\phi^\alpha_j}$ must be real and independent of $\alpha$ and $j\in Z_{1, \dimv}$. 
If $e^{-i2\phi^\alpha_j} = -1$, then $\gamma=1$.
But by the Lemma proven previously, this implies that $q=0$ and $\epsilon=1$.
We can exclude this case if we require the regulator $q$ to be non-zero.
Thus, there appears to be just one mixed-min insensitive orbit,
the one found in Section \ref{sec:main} for the real-$\vecalp{v}$ ansatz.

\end{appendix}

\end{document}